\documentclass{elsart}
\usepackage{psfig}

\begin{document}
\begin{frontmatter}

\title{Angular Resolution of the Pachmarhi Array of \v Cerenkov Telescopes}
\author{P. Majumdar, B. S. Acharya, P. N. Bhat, V. R. Chitnis,}
\author{M. A. Rahman, B. B. Singh and P. R. Vishwanath}   
\address{Tata Institute of Fundamental Research,
 Homi Bhabha Road,
 Mumbai 400 005, India.}

\begin{abstract}
 The Pachmarhi Array of \v Cerenkov Telescopes consists of a distributed
array of 25 telescopes  that are
used to sample the atmospheric \v Cerenkov Photon showers.
Each telescope consists of 7 parabolic mirrors each viewed by a single photo-multiplier
tube. Reconstruction of photon showers are carried out using fast timing information
on the arrival of pulses at each PMT.  
The  shower front is fitted to a plane and the 
direction of arrival of primary particle initiating the shower is obtained. 
The error in the determination of the  arrival direction of the primary
has been estimated using the {\it split} array method. It is found to be 
$\sim 2.4^\prime ~$ for primaries of energy $ > 3 ~TeV$. The dependence of the
angular resolution on the separation between the telescopes and the number of detectors
are also obtained from the data.  
\end{abstract}
\end{frontmatter}

\section{Introduction}

The $\gamma -$ray sky has been successfully 
observed by satellite based missions in
the energy range between 50 $keV$ and 30 $GeV$
\cite{thm95,har99}. At higher energies the steeply falling $\gamma -$ray 
spectrum makes it impossible to observe  by satellite based
detectors, primarily because of the limited detector size and exposure time 
available for a given source.
Alternatively, the ground based atmospheric \v Cerenkov technique has been
successfully exploited at these energies. Over the years, this technique has
proved to be the
most sensitive in studying the celestial $\gamma -$rays in the very 
high energy (VHE) range, $\sim 100~GeV$ - 50 $TeV$. 

Celestial VHE
$\gamma -$rays initiate an electromagnetic cascade in the atmosphere as
they enter the earth's atmosphere. The electrons and positrons in the cascade,
being relativistic, emit \v Cerenkov light as they propagate down the
atmosphere resulting in a faint 
flash of bluish light lasting a few nanoseconds. This directional
\v Cerenkov flash is detected on the ground by conventional optical light
detectors during moon-less clear nights. Taking advantage of the fact that 
the \v Cerenkov flash is highly collimated, within a cone of half angle $\sim 1^\circ $, along the direction of the 
incident particle most experiments simply limit the optical field of view 
of the \v Cerenkov telescopes to a small region of the sky.

However the main drawback of this technique is the presence of
abundant cosmic ray background which severely limit the
sensitivity of this technique.  Numerous efforts to develop methods 
to distinguish the \v Cerenkov light pool produced by cosmic $\gamma
-$rays from that by the cosmic rays led to two important techniques
based on complementary ways of viewing the cascade, {\it viz.} the 
angular sampling or the {\it imaging technique} and spatial sampling or the 
{\it wavefront sampling technique}.  
Both these techniques are currently being employed by different groups
\cite{cr93,sc96,fe97,on98}. 

The possibility of timing the \v Cerenkov light front for the estimation of 
arrival direction of the primary has been realized as 
early as 1968 by Tornabene and Cusimano \cite{tnc68}. It was later demonstrated
by the Durham group \cite{dhk84} that while viewing Crab pulsar the on-axis 
events were richer in $\gamma $-rays than off-axis events. It was also used by 
Gupta {\it et~al.}\cite{grt85} to improve the signal to noise ratio of 
atmospheric 
\v Cerenkov telescopes consisting of large mirrors with poor optical quality.
The basic principle of all these experiments is the fast timing technique 
using spatially separated \v Cerenkov telescopes.

The signal to noise ratio in such an experiment is given by~\cite{Ach93}
\begin{equation}
 {{S}\over{N}} \sim \sqrt{{AT} \over {(1-f_r)F_{c}\omega}}f_\gamma~F_\gamma
\end{equation}
where {\it $F_c$}  and {\it $F_\gamma$} are the fluxes  of omni-directional
cosmic rays  and $\gamma-$ rays from a point source respectively. {\it A} is
the collection area of the array, {\it T} is the time of observations,
{\it $\omega$} is the solid angle of acceptance, {\it $f_r$} is the fraction of
showers due to cosmic rays that are identified and rejected as background and
{\it $f_\gamma$} is the fraction of showers due to $\gamma -$rays that are
identified as signal and hence retained.
In order to achieve high ${S}\over{N}$ apart from increasing {$f_\gamma$} and
{\it $f_r$} one 
could either increase the collection area and the observation time or decrease 
the solid angle of acceptance. For a given exposure time and available hardware
resources one can 
possibly increase {\it S/N} by only reducing $\omega$ as celestial $\gamma-$ 
rays from point sources are directional while cosmic rays are isotropic. 
Due to the finite opening angle of the \v Cerenkov cone and the spread in 
the arrival angle of \v Cerenkov photons the aperture of the telescopes 
have to be restricted to few degrees, which sets a  lower limit to
{\it $\omega$}. However, it is possible to improve the ${S}\over{N}$
for point sources without losing \v Cerenkov light if 
the direction of arrival of primary particles is estimated accurately 
\cite{tnc68,grt85,ths79,prv82}. The 
shower axis retains the original direction of the primary.

	To estimate the arrival direction accurately, the error in the 
reconstructed direction,
{\it i.e.} the angular resolution has to be very small. The two dominant
factors which contribute to the angular resolution are the average distance,
{\it D}, between the telescopes and {\it $\delta$t}, the uncertainty 
in the measurement of arrival time of photons at the telescopes. For example,
the angular resolution in  the zenith angle, $\theta$ for $n$ detectors is 
given by \cite{grt85}:
\begin{equation}
{\delta\theta} = {{c~\delta t} \over  {D\cos \theta}} \sqrt {2 \over n}
\end{equation}
 
Therefore,  a large number of telescopes with instrumentation to 
measure the relative arrival time of photons and separated by large distances
are needed to  reconstruct the shower front and estimate the direction of
arrival of the shower fairly accurately. On the other hand, in case of the
imaging technique the angular resolution is limited by PMT(pixel) size which is
typically of the order of a quarter degree.  
The improvement in the signal to noise ratio by restricting the angle of
acceptance to $\delta \theta $ would be: 
$$\sim \sqrt {{\omega}\over{\pi \delta \theta^{2}}}$$  
This is a very significant advantage of non-imaging arrays with high angular 
resolution. Moreover, this method of reducing background is independent of 
the primary species. Hence it is extremely useful at very low primary energies
where the cosmic ray electrons form a significant source of background.
Electrons cannot be discriminated easily, unlike hadrons, since they too
undergo electromagnetic interactions in the atmosphere akin to $\gamma-$ rays.
However one has to bear in mind that the angular resolution could be poorer at
lower primary energies due to paucity of \v Cerenkov photons at the ground
level.   

In the rest of the paper we discuss the method of obtaining the
arrival direction of the primary radiation using our array of \v Cerenkov 
telescopes and estimate its error.

\section{\bf Pachmarhi Array of \v Cerenkov Telescopes}

The experiment at Pachmarhi (longitude:
78$^{\circ}$ 26$^{\prime}$ E, latitude: 22$^{\circ}$ 28$^{\prime} N$ and
altitude: 1075 $m$), is based on the wavefront sampling
technique and employs an array of \v Cerenkov telescopes, called the 
Pachmarhi Array of \v Cerenkov Telescopes(PACT), to sample the 
\v Cerenkov light pool. PACT is now fully functional.

 The experimental set-up of PACT has been explained in detail elsewhere
~\cite{PACT}. Briefly, it consists of a $5\times 5$ array of atmospheric
\v Cerenkov telescopes deployed in the form of a rectangular matrix with a
separation of 25 {\it m} in the N-S direction and 20 {\it m} in the E-W
direction.  Each telescope consists of 7 parabolic mirrors of 0.9 {\it m}
diameter mounted paraxially and having a focal length of 90 {\it cm}. Each
mirror is viewed by a fast photo-multiplier tube (PMT, EMI 9807B) at the focus
behind a circular mask of $\sim$ 3$^\circ$ diameter. However the field of view
is limited by the diameter of the PMT photo-cathode to $\sim$ 2.9$^\circ$
{\it FWHM}. 
  
The array has been divided into 4 sectors with six telescopes in each.
The pulses from 7 PMTs in a telescope are added linearly 
to form a telescope sum pulse called {\it royal sum}. Each
{\it royal sum} from the 6 telescopes in a sector are suitably discriminated 
(typical {\it royal sum} rates $\sim$ 30-50 kHz.) and a trigger
is generated by a coincidence of any 4 of these 6 {\it royal sums}. 
 The typical event rate is $\sim$ 2-5 Hz.  For every event, information
on  the relative arrival times and density of \v Cerenkov photons  
are recorded for the 6 peripheral mirrors/PMT in each telescope in each sector.
The relative arrival times of {\it royal sum} pulses are recorded  both 
in the respective sector and in the central data processing station.

Thus, PACT measures the arrival time of shower front at 
various locations within the \v Cerenkov light pool at two distance scales, 
{\it short range} (intra-telescope) and {\it long range} (inter-telescope). 
The arrival direction of the shower is estimated from the measured 
arrival times of \v Cerenkov photon front at 
each of the spatially separated telescopes while the six adjacent 
measurements in a given telescope could be used to study the fluctuations 
in the measured shower parameters. 
The dispersion of photon arrival times at each 
telescope contain the signature of the primary species \cite{VRC99} and 
hence could be used to distinguish between $\gamma-$ rays from the 
background \cite{VRC2k1}. The density measurements on the other hand 
enable us to estimate the energy of the primary species as well as 
to reject cosmic ray background \cite{VRC2k1b}.  

\section{\bf Alignment of Mirrors and Telescope }

\subsection{\bf Pointing Accuracy of Telescopes}
  The telescopes are equatorially mounted and each telescope is independently
steerable in right ascension and declination within $\pm 45^\circ$.
The movement of the telescopes is remotely
controlled by Automatic Computerized Telescope Orientation System
(ACTOS)~\cite{ACTOS}. The hardware for ACTOS consists of a semi-intelligent
closed loop stepper motor drive system
which senses the angular position using a gravity based angle transducer called
{\it clinometer} with an accuracy of 1$^\prime$.
The two clinometers,  one each in N-S and 
E-W direction, are
accurately calibrated using stars at various hour-angles and declinations. 

In order to estimate the source pointing error of the system, the telescopes 
were oriented to different bright stars at random positions in the sky
and the offsets in orthogonal directions (ascension and declination) with 
respect to the star were noted. Using this data it is concluded that
the system can orient the telescopes to the putative source
with a mean offset of $0^\circ.003 \pm 0^\circ.1$. 
The source pointing is
monitored with an accuracy of $0^\circ.05$ and corrected in real time during 
tracking.
The uncertainty in pointing  
represents the subjective error in manual reading of the star position at the
cross wire of the guiding telescope field of view~($\pm~0^\circ.05$).  

\subsection{Measurement of Alignment Accuracy of Mirrors}
         As the seven mirrors of a telescope are mounted on a 
single mount it is necessary to ensure that
all their optic axes are parallel to each other so that they
view the same part of the sky.
This alignment of mirrors is done manually by positioning the star image
at the centre of the focal plane.

\begin{figure}
\centerline{\psfig{file=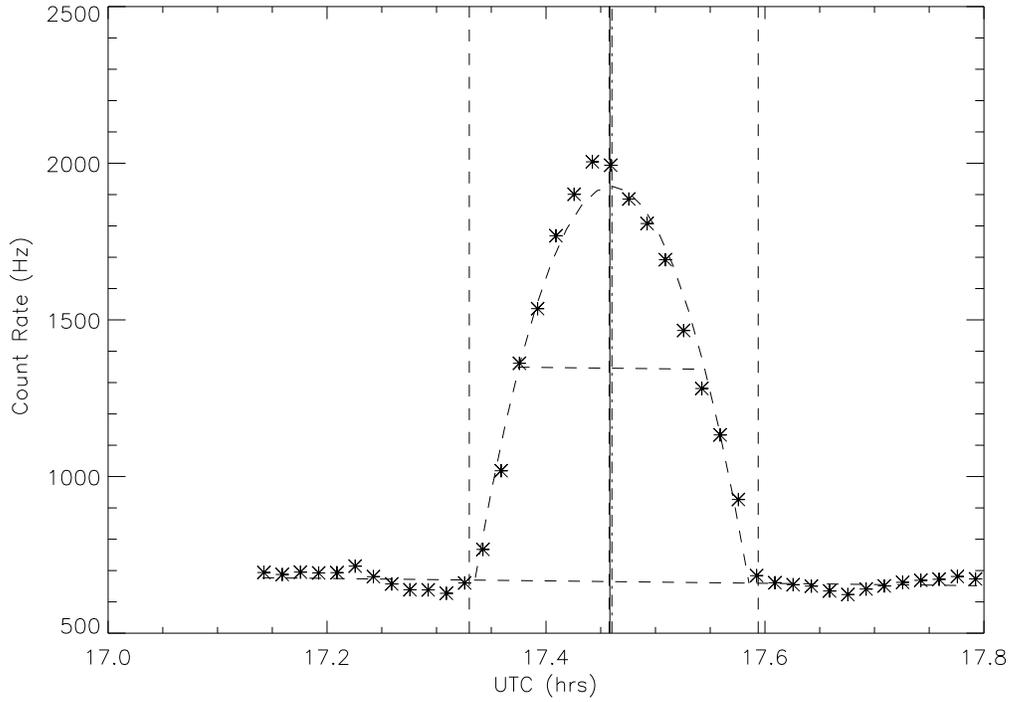,height=10cm}}
\caption{Bright Star Scan Results on $\psi-$ Ursa Major: The counting rates are 
shown as a function of time (UTC). The vertical lines around the peak of the profile 
are (i) expected transit time
(solid line) (ii) centroid of the count rate profile (dot dashed line) defined 
by two dashed lines on either side. The baseline and the FWHM are also shown.
\label{Fig:BSCAN}}
\end{figure}

   To check the accuracy of alignment of the mirrors and telescopes a
bright star drift scan is carried out. The telescope is aligned to an
isolated bright star (typically of visual magnitude 2 to 3). Then the telescope
is moved to the west by $4^\circ$ and at a suitable time the telescope tracking
is switched off.  The counting rates from each of the PMTs are monitored every
second and recorded. The count rates stay reasonably constant until the star
walks into the field of view when they increase, go through a maximum and
return to the background value as shown in figure~\ref{Fig:BSCAN}.
The background count rates before and after the star transit are fitted to a
linear function. The background is subtracted from the count rates during the
star transit by interpolation. The background corrected count rates are then
fitted to a quadratic function. Figure \ref{Fig:BSCAN} shows one such fit to a
typical count rate profile of a mirror.
Table~\ref{TAB:BSCAN} shows the summary of results of a typical bright star
scan. It shows in column 3 the FWHM of the drift scan profile of the count rates
due to the
star. The offset of the centroid of this profile with respect to 
the centre of the field of view ({\it i.e.} expected transit time) is shown in 
the last
column. This offset is the error in the alignment in right ascension. 
The relative FWHM's of the profiles could indicate the misalignment in 
declination, if any. On the other hand the absolute values of FWHM of the 
count rate profiles
are a function of the PMT gains, image quality
(point source image size $\le~1^\circ$), star brightness $etc$.
Using this method, it is ensured that the optical axes of all
the 7 mirrors in a telescope are parallel to each other within an error of
$0^\circ.2$ or less. If the error exceeds this value the particular 
mirror is re-aligned and re-checked. This method of alignment is similar to
that adapted for aligning the imaging telescope arrays \cite{kaa96}.

\begin{table} \centering
\caption {Bright Star Scan Results of $\psi-$ Ursa Major for a telescope
             \label{TAB:BSCAN} }
\vskip 0.25 cm
\begin{tabular}{|l|l|l|l|}
\hline

Mirror  &  Peak Count rate ($Hz$) & FWHM (deg.) & Offset (deg.)  \\ \hline

\hskip 0.4 cm A   & \hskip 0.6 cm   9430   & \hskip 0.4 cm  2.29  & \hskip 0.4 cm -0.23      \\ \hline
\hskip 0.4 cm B   & \hskip 0.6 cm   1260   & \hskip 0.4 cm  2.46  & \hskip 0.4 cm  0.03       \\ \hline
\hskip 0.4 cm C   & \hskip 0.6 cm    847   & \hskip 0.4 cm  2.50  & \hskip 0.4 cm -0.14      \\ \hline
\hskip 0.4 cm D   & \hskip 0.6 cm    781   & \hskip 0.4 cm  2.80  & \hskip 0.4 cm  0.09       \\ \hline
\hskip 0.4 cm E   & \hskip 0.6 cm    493   & \hskip 0.4 cm  2.67  & \hskip 0.4 cm -0.07       \\ \hline
\hskip 0.4 cm F   &  \hskip 0.6 cm   616   & \hskip 0.4 cm  2.18  & \hskip 0.4 cm -0.19      \\ \hline
\hskip 0.4 cm G   & \hskip 0.6 cm   2070   & \hskip 0.4 cm  2.38  & \hskip 0.4 cm  0.24        \\ \hline

\end{tabular}
\end{table}

\section{Estimation of Timing Resolution}
      In order to estimate the angular resolution one has to estimate the error
in timing measurement. To determine the arrival time of photons accurately, the
PMT should have a fast response time and low dispersion in the photo-electron
transit times.  We use a low noise, high gain tube having rise-time of
$\sim$ 2 {\it ns}, pulse width of 3 {\it ns} and transit time jitter of
1.9 {\it ns} (FWHM). The PMT signals are brought to the recording
stations through low loss coaxial cables (RG 213). The intrinsic timing jitter 
of the signals from the PMTs limit the accuracy of timing measurements.

The timing resolution is estimated from the data. The relative
delays of 6 peripheral PMT's of each telescope and {\it royal sum} pulses with
respect to the 
trigger pulse were recorded  using the fast time to digital 
converters(TDC). The signals were delayed suitably using ECL based 
delay generators having an intrinsic stability of $\pm 0.1 ns$. 
Linearity of TDC was monitored by periodically calibrating them with standard
cable delays. The LeCroy(Model \# 2228A) and Philips Scientific(Model \# 7186)
TDC modules that are used in the experiment are set to a sensitivity of 
0.25 and 0.2 {\it ns} per count, respectively. 

Data are collected with all telescopes pointing to zenith. The width of the
distribution of differences in arrival times of signals ($\delta t_{ij}$) of two
TDC channels is an indication of the limiting accuracy of timing measurement,
provided the signals originate from PMT's located nearby. The distribution of
{\it $\delta t_{ij}$} tends to broaden due to one or more reasons mentioned below:
\begin{description}
\item {(a)}  fluctuations in the arrival time and density of \v Cerenkov photons.
\item {(b)} errors in mirror  pointing result in sampling different 
  angular regions, which in turn leads to pulse slewing. 
\item {(c)} spatial separation between the mirrors/telescopes. The mean arrival
time as well as its fluctuations is a function of core distance~\cite{VRC99}. 
\item {(d)} fluctuations in electron transit time in PMT's,  the differences in
the cable delays and the propagation delays in the electronic circuits. 
\end{description}

    Delays due to unequal cable lengths, and propagation delays in the 
electronic circuits have been matched to an accuracy of about 0.25 ns. 
Mirrors are aligned to within $\pm~0^\circ.2$ as mentioned earlier. The effect
of slewing responses of the PMT's is often determined by using filter wheels
to attenuate laser calibration light flashes transported to each mirror through
optical fibers \cite{obb01}. No such measurements were made in this 
case since the pulse rise time is still reasonably small ($\le~2.4~ns$). The 
effects due to finite separation between the detectors is corrected as
discussed below and shown in figure 2.

In the case of signals from individual PMT's of a telescope 
the distances between the detectors is of the order of a metre. 
Only those combinations corresponding to neighboring PMT's are 
considered in order to minimize the contributions from fluctuations in the 
arrival time of \v Cerenkov photons and distance dependent effects, 
to the uncertainty in timing measurement. 
The results are shown in the table~\ref{TAB:sigmaij}. 
The limiting accuracy of timing measurement is about 1 $ns$ for individual 
PMT signals.   

\begin{figure}[hbt]
\centerline{\psfig{file=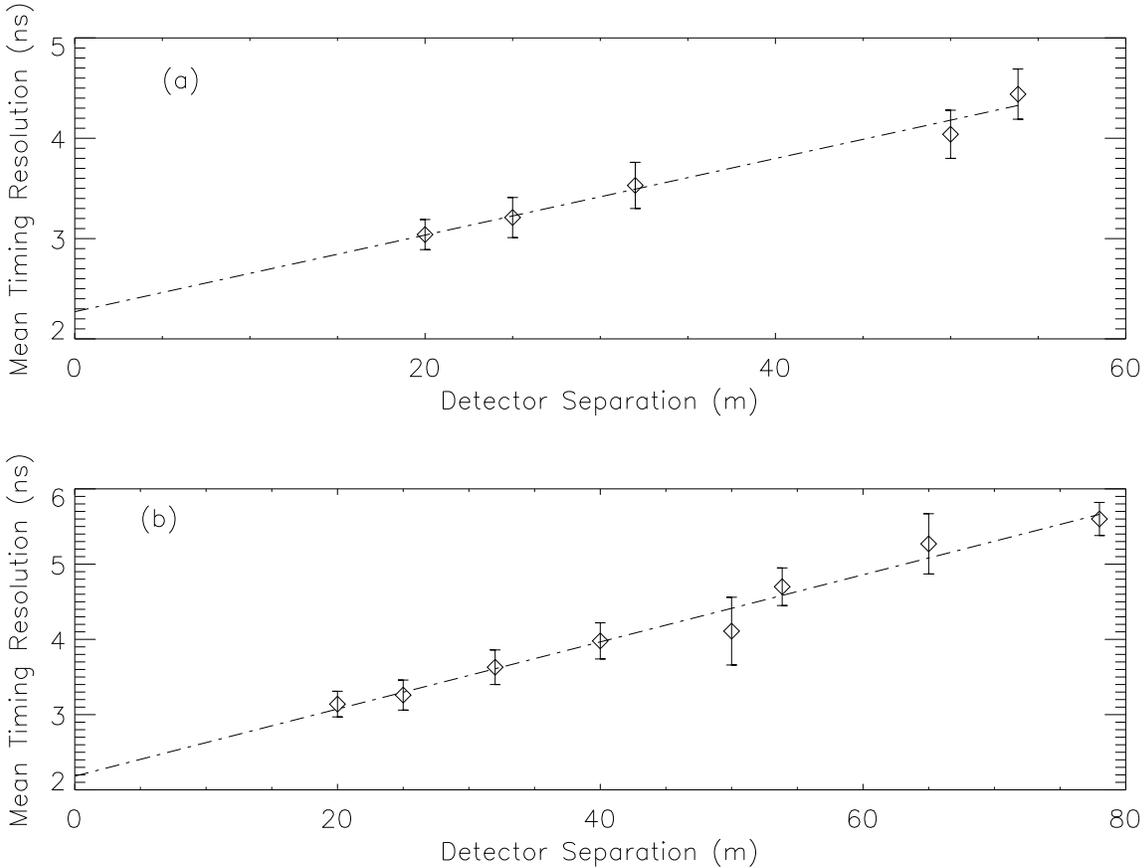,height=12cm}}
\caption{Figure shows the variation of $<\sigma_{ij}>/\sqrt{2}$, 
derived from the TDC difference distributions of {\it royal sum} pulses, as a 
function of detector separation for (a) single sector (\# 3) and 
(b) for 2 sectors. Fitted straight lines extrapolated to zero separation 
yield the required timing resolution independent of telescope separation. 
\label{Fig:sigmaij}}
\end{figure}

In the case of {\it royal sum} pulses, the distance between neighboring
telescopes is 20 {\it m} or more, which is too large to be ignored. Moreover, 
the spread in the arrival time depends upon how well the signals of a
telescope are added.  To remove the distance dependent effects,
the standard deviation $\sigma_{ij}$ are plotted as a function 
of separation between the telescopes, $i$ and $j$ and is shown in 
figure~\ref{Fig:sigmaij}.  In the past, the latter effect was
minimized by tilting the telescopes towards each other by about a degree
and the timing resolution was measured \cite{tnc68}. However we feel that the 
present method of reducing contribution 
from the detector separation is more accurate. From the figure it can be seen
that the error $\sigma_{ij}$ clearly increases with the separation as 
expected. The FWHM for two telescopes side by side is estimated by 
extrapolation, thus removing the effect of finite separation between the 
telescopes on the  $\sigma_{ij}$ distribution.
The corresponding values are shown in table~\ref{TAB:sigmaij}. These values are
comparable to the intrinsic timing jitter for proton primaries estimated from 
simulation studies for PACT telescopes \cite{VRC2k1}. This means that the 
contribution to timing resolution from instrumental effects is negligibly small.

\begin{table} \centering
\caption{Timing Resolution of PACT Array \label{TAB:sigmaij}}

\vskip 0.25cm
\begin{tabular}{|l|l|l|}
\hline

\hskip 0.3 cm Timing      & \hskip 0.3 cm Sector                      & Timing \\
\hskip 0.3 cm Information & \hskip 0.5 cm \#  & Resolution($\sigma_{i}$~{\it ns}) \\
\hline

           {\it Royal Sum}   & \hskip 0.7 cm  3  &   2.3$\pm$0.1  \\ \hline
           {\it Royal Sum}   & \hskip 0.7 cm  4  &   2.1$\pm$0.2  \\\hline
           {\it Royal Sum}   & \hskip 0.4 cm 3+4 &   2.2$\pm$0.1  \\ \hline
           {\it Individual PMT}  & \hskip 0.7 cm  3  &   1.1$\pm$0.1   \\ \hline
           {\it Individual PMT}  & \hskip 0.7 cm  4  &   0.8$\pm$0.1  \\

\hline
\end{tabular}
\end{table}

\section{Estimation of Arrival Direction of Shower}

 The arrival direction of a shower is determined by 
reconstructing the shower front using the relative 
arrival time of \v Cerenkov photons at each telescopes.
We fit the \v Cerenkov photon  
front to a plane using the measured arrival times, the 
normal to this plane gives the direction of shower axis. 

\subsection{Calculation of Time-offsets}

   The relative arrival time of pulses as measured in the experiment is not
the relative arrival time of \v Cerenkov photons at the PMT, needed
for reconstructing the shower front.
A finite but constant delay between pulses from different PMT's 
arise due to unequal cable lengths, differences in electronic propagation
delays and differences in photo-multiplier transit time {\it etc}. These are termed as
T0 or time-offsets. Thus the measured relative arrival times have to be
corrected for this time-offsets to get the relative arrival time of
\v Cerenkov photons at the PMT. In the absence of a calibration light source
which could generate pulses in all PMT's simultaneously,
these time-offsets have been  determined from the data itself~\cite{sinha}.

The time difference between \v Cerenkov photons arriving at PMT's
(or telescopes) located side by side should be zero on the average for vertical
showers. Thus
the average of time delays between two PMT's, from a large sample of data,
is entirely due to difference between the two time-offsets. 
We chose vertical showers in order
to eliminate the errors in the estimation of offsets due to finite telescope
separations for inclined showers.

If $T0_i$ and $T0_j$ are the time offsets for the PMTs {\it i} and {\it j}, we can
write an equation of the form
\begin{equation}
 (T0_i - T0_j)  = C_{ij}
\end{equation}

where {\it $C_{ij}$} is the mean delay between a pair of PMT's {\it i} and {\it j} after
correcting for the time difference due to differences in height
(z-coordinates) of PMT's(or telescopes) .

The total $\chi^{2}$ can be expressed as

\begin{equation}
\chi^{2}=\sum_{i,j=1; i\neq j}^{n} W_{ij} (T0_i - T0_j -  C_{ij})^{2}
\end{equation}

where $n$ is the total number of PMT's, {\it $W_{ij}$}'s are 
the statistical weight factors, 
$(W_{ij} = 1/{{\sigma^C}_{ij}}^{2}$, where ${\sigma^C}_{ij}$ is the 
uncertainty in determining {\it $C_{ij}$}). $\chi^{2}$ minimization will give a
set of $n$ equations of the form

\begin{equation}
\sum_{j=1; i\neq j}^{n} W_{ij} (T0_i - T0_j) =  \sum_{j=1; i\neq j}^{n} W_{ij} C_{ij}=C_{i}^{\prime}
\end{equation}
Thus we have a set of simultaneous equations involving {\it W}, {\it T0} and
$C^\prime$.
All these sets of equations can be written in the form of matrix equation
\begin{equation}
W~T0=C^\prime
\end{equation}
Solving this matrix equation one gets an estimate of time-offsets for {\it (n-1)}
detectors with respect to {\it $n^{th}$} detector. It is to be noted that the differences in 
time delays due to the differences  in the heights of the telescopes were 
taken off while calculating these time-offsets, as already mentioned.

However the mean time difference $<\delta t>$
({\it ns}) is a function of core distance R ({\it m}) and varies, for vertical 
showers, as
$$<\delta t> = \left( {{\delta d}\over {Hc}}\right) R$$ where $\delta d$ is the 
separation between the
telescopes, {\it H} the radius of curvature of the \v Cerenkov shower front 
and {\it c} the velocity of light. Typically,
$<\delta t> = 0.0117R~ {\it ns}$ for the energy and altitude concerning us
({\it H} $\sim$ 10 {\it km}). 
In our experiment, the average separation between the telescopes in a sector is
35 {\it m}-50 {\it m} whereas the separation between the PMT's in a telescope 
is of the order of a metre. This can introduce a systematic error in the 
estimated values of T. For a proton shower of average energy ($\sim 1.8~TeV$)
the mean collection radius is $\sim 100~m$. Using this it is estimated that
the maximum systematic error in T arising from the \v Cerenkov shower front
curvature  is $\sim 1.2~ns$.

\subsection{Reconstruction of Arrival Direction}
   Using the plane front approximation the arrival direction of the shower is 
estimated as follows \cite{sinha,Ach93}.
          If $x_i$,$y_i$,$z_i$ are the coordinates of the 
$i^{th}$ PMT, $l,m,n$ the direction cosines of the shower axis and 
$t_i$ the arrival time of the photons at this PMT then the equation relating 
them is given by   
\begin{equation}
   lx_i+my_i+nz_i+c(t_i-t_0)=0
\end{equation}

     where $t_0$ is the time at which the shower front passes through the 
origin of the coordinate system. Then the arrival direction of the shower 
can be estimated by a $\chi^{2}$ minimization where
\begin{equation}
\chi^{2}=\sum_{i=1}^{n} w_i ( lx_i +my_i +nz_i +c(t_i - t_0) )^{2}
\end{equation} 

 where $w_i$'s are the statistical weight factors for the $i^{th}$ timing 
measurement(= 1/$\sigma_{i}^2$ where $\sigma_{i}$ is the uncertainty in the timing
measurement for the $i^{th}$ detector). The values of $l,m,n$ and $t_0$ are 
calculated using the 
equations 
{$\partial\chi^{2}/\partial l = 0$}, {$\partial\chi^{2}/\partial m = 0$},
{$\partial \chi^{2} / \partial t_0 = 0$} and \(l^{2}+m^{2}+n^{2}=1\).
We have ignored terms containing $\partial n/ \partial l$ and $\partial n/ \partial m$
as these are small.
The contribution to the error in
angle determination is negligibly small up to an angle of $\sim 80^\circ$ under
this assumption.

\begin{figure}[hbt]
\centerline{\psfig{file=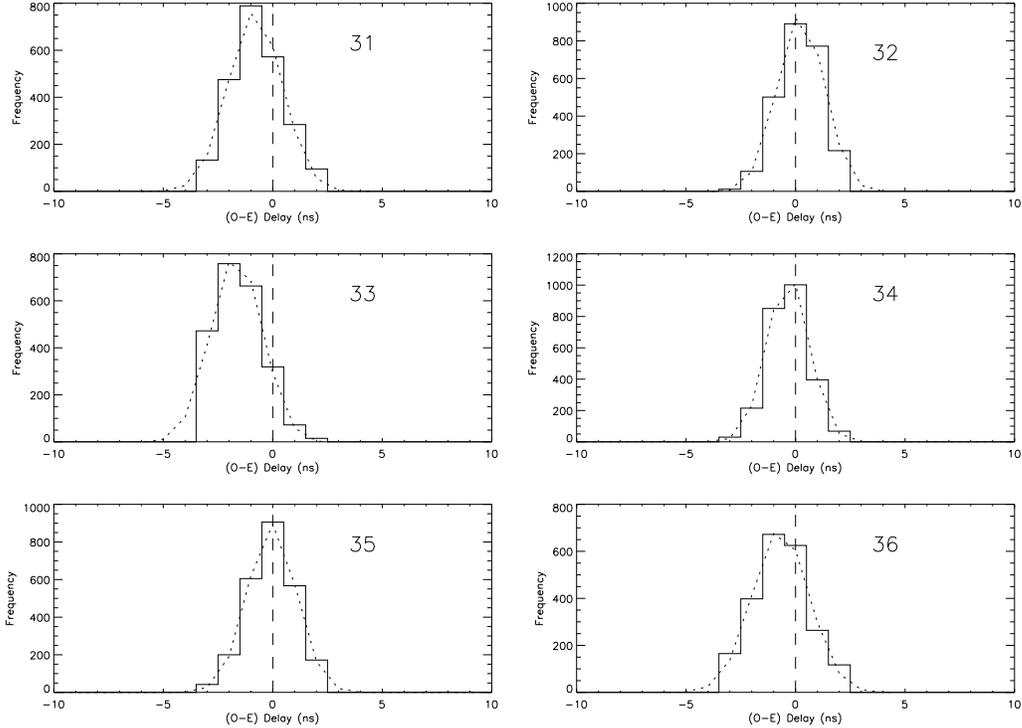,height=10cm}}
\caption{{\it Observed - Expected} Delay Distributions for various telescopes in 
a sector, based on {\it royal sum} pulses. It is seen that the $\sigma$ of the 
delay distribution is about 1 {\it ns}. \label{Fig:delay}}
\end{figure}

In the first iteration all PMT's with valid delays are
used for the fit and the  {\it observed - expected} delays are calculated for
each. In the subsequent iteration, the PMT with the largest deviation is
rejected if the absolute  difference between observed and expected delays 
is greater\footnote{$\pm 3 \sigma$, The
standard deviation of {\it observed - expected} delay $\sim~1~ ns$ }
than 3 {\it ns}. This is to minimize the effect of isolated large fluctuations
in the photon arrival times \cite{mw99}.
Figure~\ref{Fig:delay} shows the {\it observed - expected} delay distributions 
for various telescopes, based on {\it royal sum} pulses.
This process of fitting is continued until there are no more large deviations
 or the number of available PMT's have become less than 4, in which case the 
shower front reconstruction process is aborted. 

\begin{figure}[hbt]
\centerline{\psfig{file=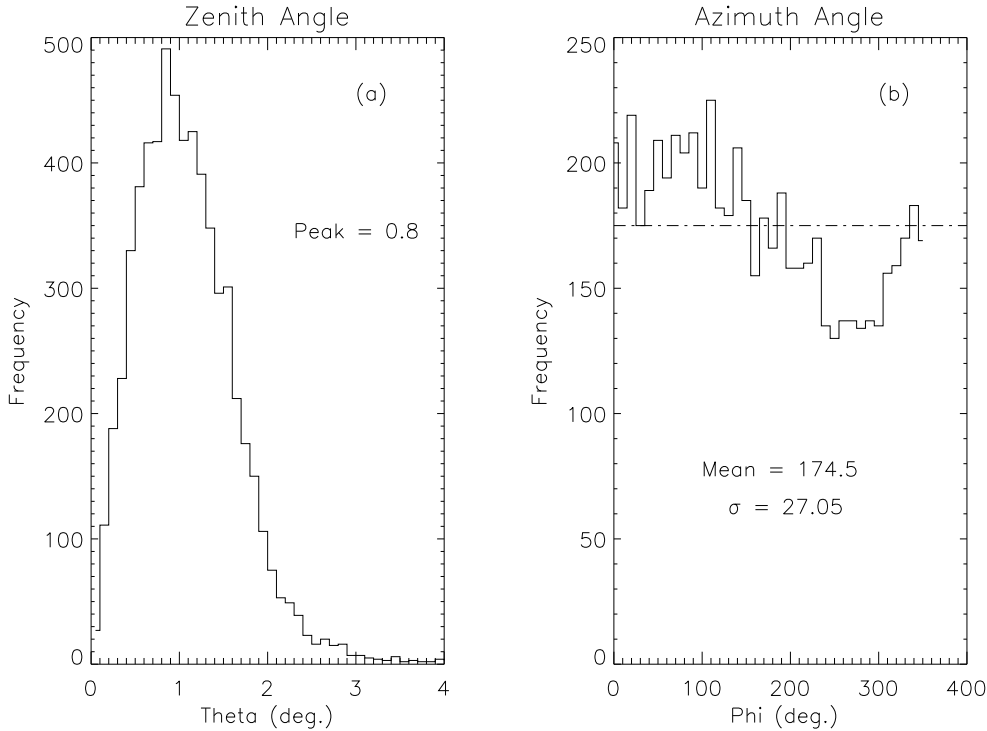,height=10cm}}
\caption{Distribution of zenith and azimuth angles estimated from $royal~sum$
timing informations from all telescopes. \label{Fig:angle}}
\end{figure}

Figure~\ref{Fig:angle} shows the zenith and azimuthal angle distributions of 
the reconstructed arrival directions using the procedure explained above. The
reconstruction of the shower front was carried out using 25 telescope data 
collected with telescopes in the  vertical position. 
The observed distributions are consistent with those expected.

      Further, it has to be ascertained that the telescopes are pointing to the
{\it true} zenith direction. A tilt in the observer's 
co-ordinate system will imply a systematic error in pointing to the source 
direction. It is ensured that the assumed horizontal plane is really 
perpendicular to the direction of {\it true} zenith from the
projected angle distributions of the fitted directions.
The projected angles $\alpha$ and $\beta$ are related to the zenith angle 
$\theta$ and azimuth $\phi$ by

$$tan~ \alpha = tan~ \theta ~ cos~ \phi$$ and
 $$tan~ \beta = tan~ \theta~ sin~ \phi$$

\begin{figure}
\centerline{\psfig{file=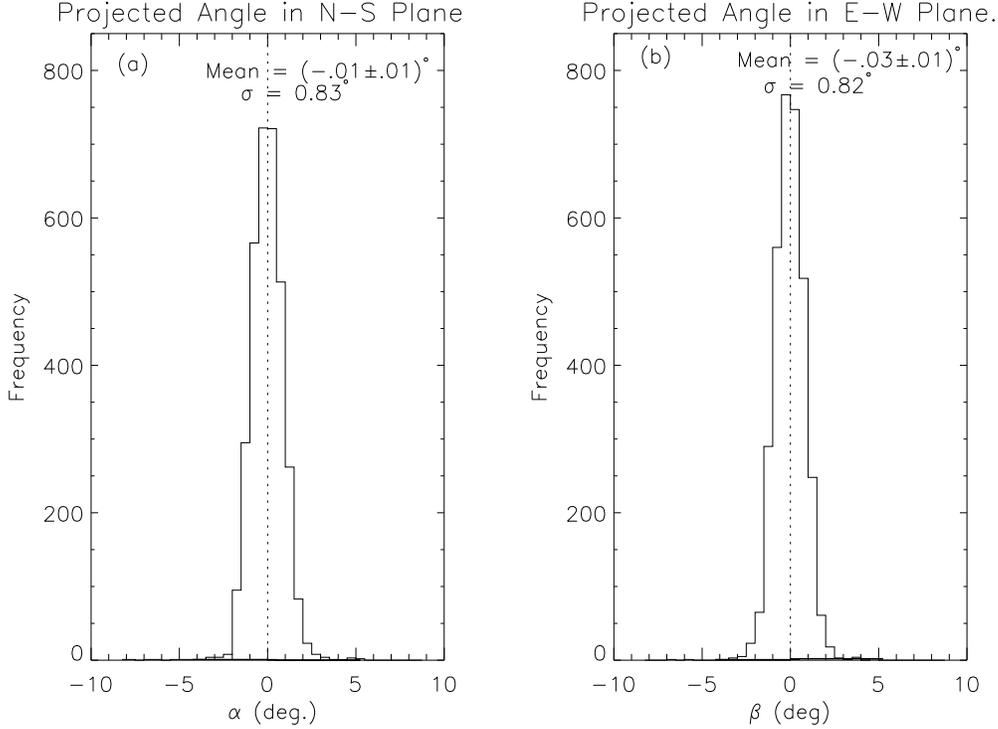,height=10cm}}
\caption{Distribution of Projected Angles, $\alpha$ and $\beta$, in the (a) N-S 
and (b) E-W planes. The dot-dashed lines are Gaussian fits to the distribution
 \label{Fig:project}}
\end{figure}

\noindent {where} $\alpha$ and $\beta$ are the projected angles in the N-S and E-W
plane respectively. The frequency distributions of $\alpha$ and $\beta$ are
 shown in figure~\ref{Fig:project}. The ratio of the number of events above
and below zero are  1.0 $\pm$ 0.03 and 0.96 $\pm$ 0.03  for $\alpha$ and $\beta$
respectively. The symmetry of these distributions thus verify the 
orthogonality of the telescope pointing direction to the horizontal.  

\section{Angular Resolution of PACT}

We define the angular resolution of the array by the angle which corresponds to
the 68\% acceptance of mainly proton events.  This definition is consistent with
that for the HESS array except that their definition refers to $\gamma $-ray
showers only \cite{ahk97}. The angular resolution of PACT has been estimated by
using the {\it split} array method. The array is divided into two independent 
parts and the arrival direction is estimated for each shower from these two
sub-arrays. The distribution of space angle ($\psi$) between these two estimates
is a measure of the angular resolution, {\it i.e.} the accuracy with which one
can estimate the arrival direction.  Since there are two independent estimates 
of the direction, the angular resolution is given by the 
peak\footnote{for small angle approximation $\sigma_\psi$ of the point spread 
function coincides with the peak of the space angle ($\psi$) distribution} of 
the distribution of space angle between the two directions 
as \(Peak \over \sqrt{2}\).

\begin{figure}
\centerline{\psfig{file=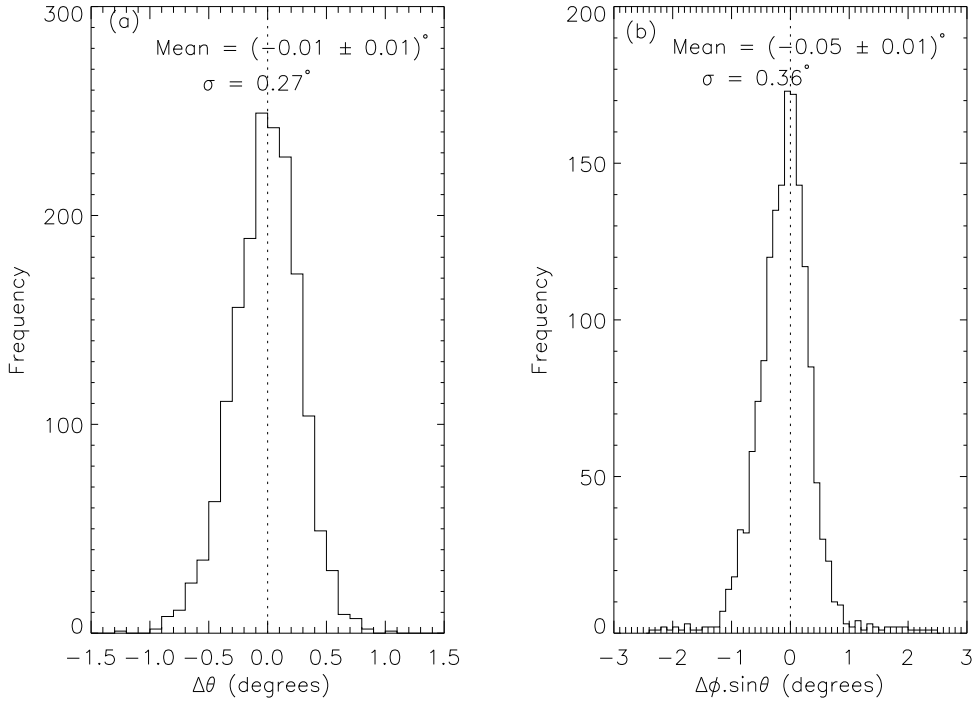,height=10cm}}
\caption{The differences of two independent estimates $viz.$ $\Delta\theta$ and 
$\Delta\phi$ distributions using 6 telescopes for each estimate. 
\label{Fig:delta}}
\end{figure}

Similarly,  the uncertainties in the zenith
angle $\theta$ and azimuth $\phi$ are estimated from the distributions of
$\Delta \theta$ and $\Delta \phi$, the differences between the corresponding 
angles from two  independent estimates for a shower. 
Figure~\ref{Fig:delta} shows these distributions for
$\Delta\theta$ and $\sin {\theta}.\Delta\phi$, respectively.

\subsection{ Using Royal Sum pulses}

\begin{figure}[hbt]
\centerline{\psfig{file=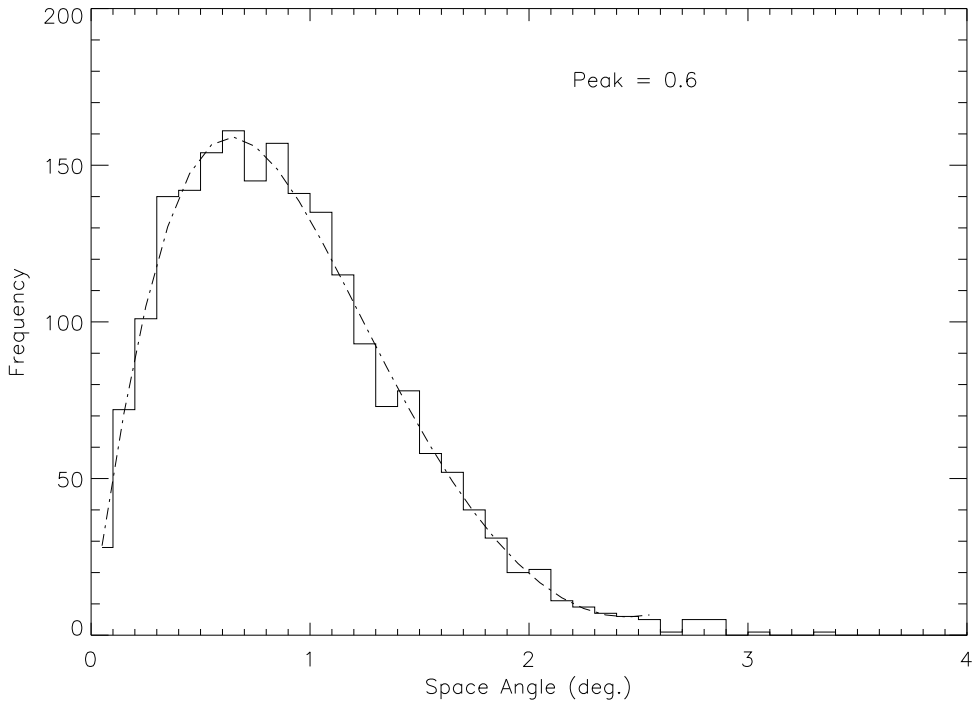,height=10cm}}
\caption{Distribution of space angle between two measured directions. The 
estimates are made with 12 telescopes from sector 1,2 and 12 from Sector 3,4.
The dot-dashed line is a fit to the observed distribution. 
\label{Fig:royal-split}}
\end{figure}

 The array is divided into two independent parts of equal number of 
telescopes, say 12 telescopes from sectors 1,2 and sectors 3,4.  
The arrival direction is estimated for each shower from these two 
independent arrays using the relative arrival times of {\it royal sum} pulses.
The distribution of the space angle between the two estimated directions
is shown in figure~\ref{Fig:royal-split}.
Several combinations were tried to understand the dependence of angular
resolution on the separation between the telescopes and/or the number of detectors.
The results are summarized in table~\ref{TAB:royal-reso}. 
The term `odd-even' refers to 3 telescopes each from Sector 3 and 4 grouped
into one set of 6 telescopes and the remaining 6 grouped into another. 

\begin{table}[hbt] \centering
\caption{Angular Resolution of PACT using Royal Sum pulses.
\label{TAB:royal-reso}}
\vskip 0.25cm

\begin{tabular}{|l|l|l|l|l|}
\hline
~~~No.of &Combination of& Average          &~~Peak of Space   &Angular    \\   
telescopes &Detectors fitted & detector &~Angle Difference&Resolution     \\  
     &         &   separation ({\it m}) &Distribution(deg.)&(deg.)   \\
\hline  
6     & Sector 3 {\it vs} 4 &      31.8      &   ~~~~~~~0.88  &           0.62         \\ \hline
6     &    Odd {\it vs} Even   &       48.8       &  ~~~~~~ 0.64  &         0.45           \\ \hline
12     & Sector 3 and 4 {\it vs}          &               &            &                          \\
      & Sector 1 and 2 &           40.1       &      ~~~~~~~0.59  &        0.42           \\ \hline
12     & Sector 1 and 3 {\it vs}           &               &              &                        \\
      & Sector 2 and 4 &         52.0       &      ~~~~~~~0.47   &        0.33            \\ \hline
\end{tabular}
\end{table}

\begin{figure}[hbt]
\centerline{\psfig{file=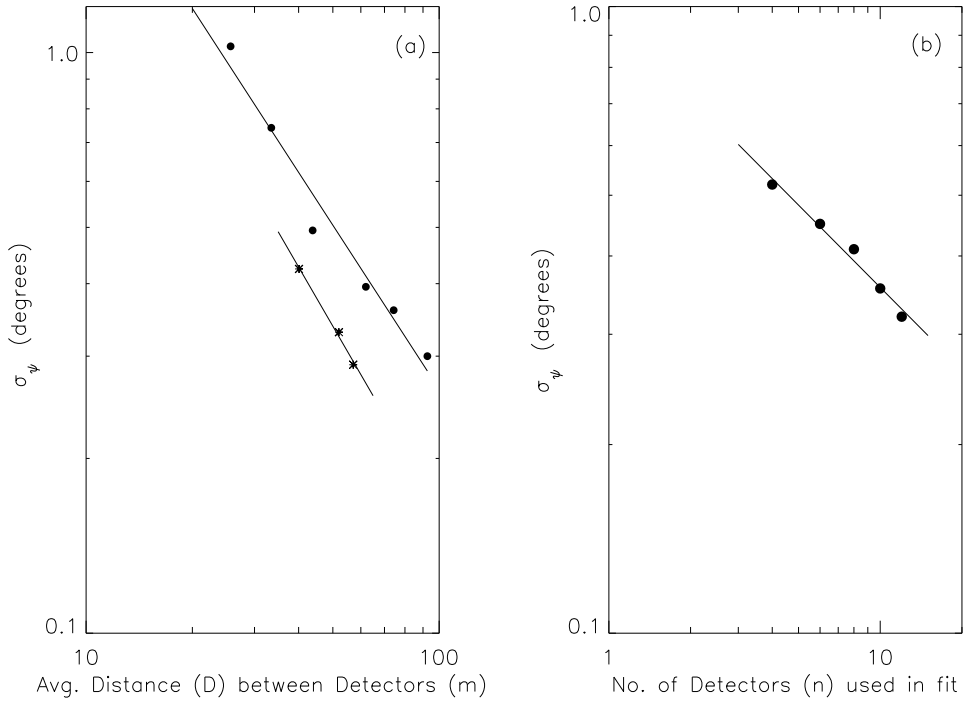,height=10cm}}
\caption{Dependence of angular resolution $\sigma _{\psi}$, as a function of 
(a) separation 
between detectors and (b) the number of telescopes. \label{Fig:dnfit}} 
\end{figure}

It is seen from table~\ref{TAB:royal-reso}  that the angular resolution  
improves as the separation between the detectors increases and also as the 
number of detectors increases, as expected.
 The dependence of angular resolution on {\it D}, the distance between telescopes 
and {\it n}, the number of telescopes used in the fit are shown in 
figure~\ref{Fig:dnfit}. 
The panel (a) of the figure shows the variation
of angular resolution with {\it D} for (i) 4 telescopes and (ii) 12 telescopes 
($\bullet$ and $\ast$ respectively). The dependence on {\it D} is calculated
to 
be $\sigma_\psi \sim D^{-0.94 \pm 0.08}$ and 
$\sigma_\psi \sim D^{-1.05 \pm 0.05}$ 
respectively, which is compliant with theoretically expected {\it 1/D} dependence.

   Once the {\it 1/D} dependence is established, the angular resolution is then corrected 
for the distance 
to extract the dependence on the number of degrees of freedom. This is shown in the panel (b)
of figure 8.  
It is seen that $\sigma_\psi \sim n^{-0.41 \pm 0.11}$, consistent with what one expects,
on a simple statistical basis, that $\sigma_\psi$ will decrease as
\(1 \over \sqrt{n}\).

It is seen from table~\ref{TAB:royal-reso} that the  best value  
of angular resolution, obtained using 12 {\it royal sum} pulse 
information, is $0^\circ.33$.  
Since we have  25 telescopes, there will be a further 
improvement in the  angular resolution by a factor of $\sqrt{2}$ 
when the arrival direction is estimated using all the telescopes in the array.  
Thus the angular resolution of the array is estimated to be  
$0^\circ.24$, from the {\it royal sum} information of all telescopes. 
The corresponding 
threshold energy of the primary $\gamma-$ray is about 1.5 $TeV$.

\subsection{Using Individual PMT timing pulses}

\begin{figure}
\centerline{\psfig{file=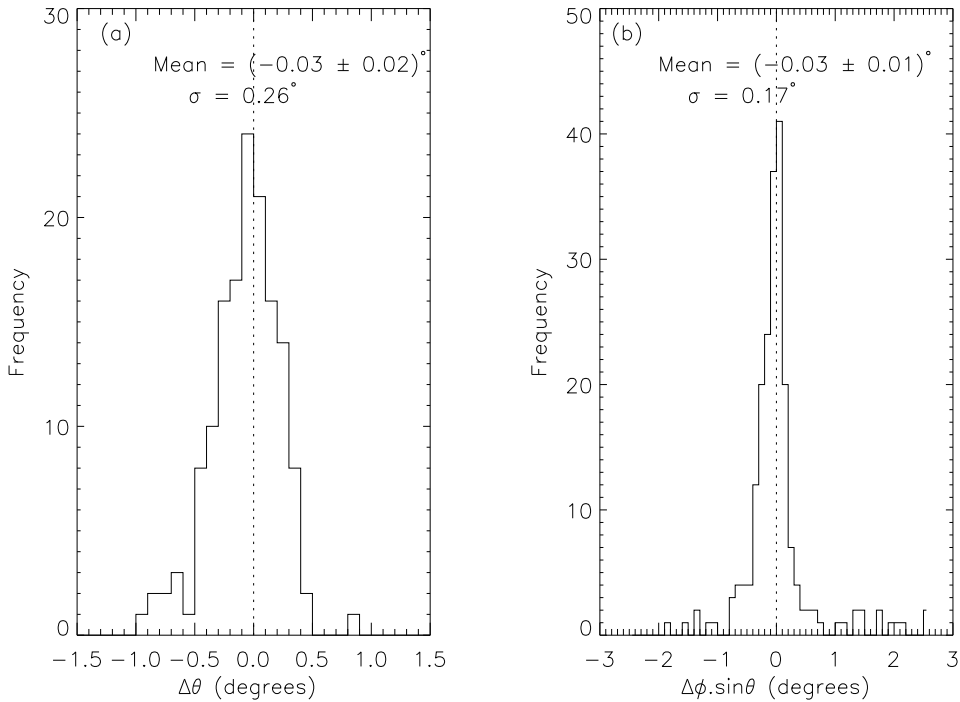,height=10cm}}
\caption{The differences of two independent estimates $viz.$ $\Delta\theta$ and
$\sin {\theta}.\Delta\phi$ distributions using individual mirrors of sector 3.
\label{Fig:cross}}

\end{figure}

 The information on the relative arrival times of \v Cerenkov photons at 
the individual PMT's are available only within a sector. Only those events 
with information in all 6 telescopes in sectors 3 and/or 4 are used 
for estimating angular resolution. 
It is obtained by dividing the data from the sectors 
into 2 subsets and obtaining the space angle between the two arrival 
direction estimates from the two subsets of data. The distributions of 
deviations of
zenith and azimuth angle estimates are shown in figure~\ref{Fig:cross} which
corresponds to data from  sector 3.

\begin{figure}
\centerline{\psfig{file=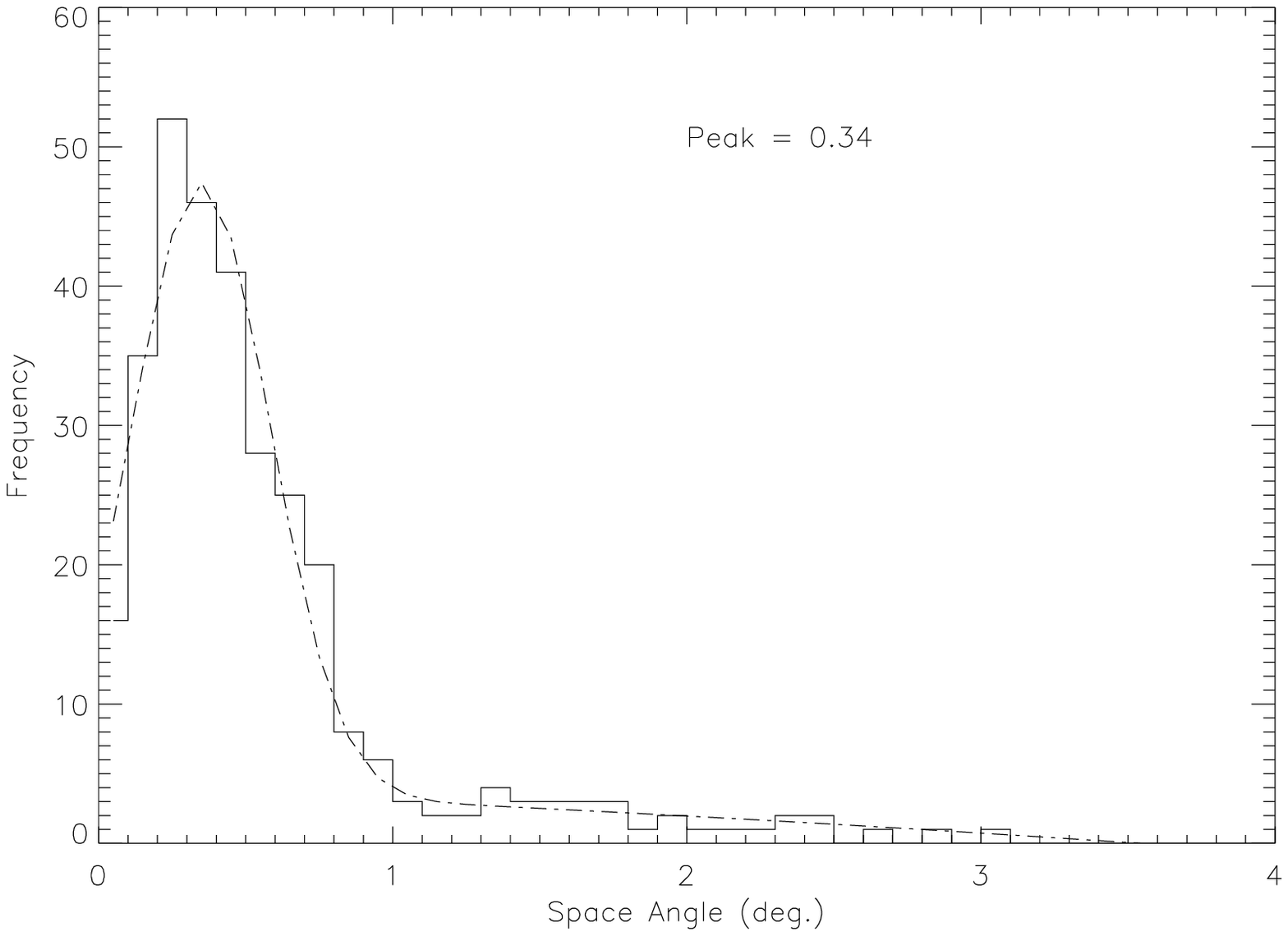,height=10cm}}
\caption{Distribution of space angle obtained from Sector 4 data,
 at least 2 PMT  in a telescope have to have valid TDC \label{Fig:s4pmt}}
\end{figure}

Information for the 3 PMT's(labelled {\it A}, {\it C} and {\it E}) are grouped into one set 
while those for the remaining 3 PMT's of a telescope({\it B}, {\it D} and {\it F}) are grouped 
into another set. The results are summarized in table~\ref{TAB:PMT-reso} . 
The rows 1 and 4 correspond to the cases in which
two subsets are obtained by demanding all {\it A} PMT's of 6 telescopes, all 
{\it B} PMT's etc. Rows 2 and 5 
correspond to cases in which the 1st group consists of at least one valid 
TDC information in any of 3 {\it A}, {\it C}, {\it E} PMT' and the second group from any one 
of 3 {\it B}, {\it D}, {\it F} PMT's. Similarly the rows 3 and 6 correspond to two sets with 
at least 2 valid TDC's in each telescope. Finally, row 7 refers to the
case in which the arrival directions are obtained separately from sector 3 
and sector 4 events and collating event arrival times to pick common events. 
Column 2 shows the  corresponding number of detectors used in the fit for 
all cases. Figure~\ref{Fig:s4pmt} shows the distribution of space angle between
the two direction estimates made using individual PMT signals in a sector.

\begin{table}[hbt] \centering
\caption{Angular Resolution of PACT using Individual PMT Information
\label{TAB:PMT-reso}}
\vskip 0.25cm
\begin{tabular}{|l|l|l|l|l|}
\hline

Sector  & No. of Detectors & Combination of  & Peak of Space & Angular \\
  \#       & used in the fit            &  Detectors      & Angle Distri- &  Resol-  \\
           &                            &                 & bution (deg)  &  ution (deg.) \\
\hline

3        &  6         &   all A, all B, etc   &  0.46           &   0.33   \\ \hline  
3        &   $\geq$ 6         &  at least 1 in &  0.45           &   0.32        \\
          &                  &  each telescope &                   &                  \\ \hline
 3        &$\geq$ 12    &  $\sim$2 in a telescope & 0.43           &  0.3          \\ \hline
 4       &      6       &  all A, all B, etc         &    0.48   &  0.34     \\ \hline
  4       & $\geq$ 6    &   at least 1 in &                       &                   \\
          &              &     each telescope     &     0.39         &  0.28                 \\ \hline
 4       & $\geq$ 12   &   $\sim$2 in a telescope & 0.34         &    0.24          \\ \hline
3 and 4   &  $\geq$ 25   &   $\sim$~5 in a telescope  &   0.2       &    0.14              \\ \hline 
                                   
\end{tabular}
\end{table}

     A better estimate of arrival direction of the shower is possible with the
use of individual PMT signals as seen in table~\ref{TAB:PMT-reso}. 
The angular resolution obtained using individual PMT's of just one sector
is $0^\circ.24$ (row 6 of table \ref{TAB:PMT-reso}).  When one considers the 
entire array the average separation doubles while the number of degrees of
freedom quadruples.  Consequently,
there will be an improvement by a factor 4 in the accuracy of the arrival 
angle estimate. Based on these considerations, the angular resolution of the
entire array is estimated to be, {$\sigma_\psi$ $\sim$  $0^\circ.06.$}, for
$\ge$ 2 active PMT's per telescope\footnote{It may be noted that the average number of timing
signals from individual mirrors in a sector is 15 when we demand $\ge$ 2 PMT's
and 27 when we demand $\ge$ 4 PMT's in each telescope.}. 

We have 6 timing information from each telescope. If we use the timing 
information from a totality of 150  PMT's in the entire array, then the angular
resolution works out to be:
$$ {\sigma_{\psi} \sim  0^\circ.04}$$
The same figure is arrived at when we use row 7 of table  \ref{TAB:PMT-reso}.
This estimate is based on the fact that $\ge$ 4 PMTs have valid TDC information
in a telescope.  The primary energy
threshold for such events is relatively higher, about 3 $TeV$. The errors in
the arrival direction estimate are larger for lower energy showers.
While the angular resolution of the imaging telescopes are limited by the 
PMT sizes, only the future imaging telescope arrays 
(like VERITAS \cite{bil99} or HESS \cite{aha97}) could achieve a better 
angular resolution. PACT is able to achieve this because it is a 
distributed array with multiple sampling of \v Cerenkov light pool. 

\section{Discussions}

For a wavefront sampling experiments such as  PACT, the technique by which 
off-axis events are rejected efficiently using their fine angular resolution, 
is vital. Hence significant effort has gone in to the improvement of the
angular resolution of such 
arrays. In the case of the imaging telescopes the angular resolution is mainly
decided by the size of the photo-tubes in the imaging camera at the focal point.
However a new technique of stereoscopic viewing has been developed in 1996 by
HEGRA group wherein the same shower is imaged by multiple telescopes. Using this
technique
the shower axis can be completely reconstructed in space using the information
concerning the location of the images within each cameras and
their angular orientation \cite{kaa96}. The reconstruction procedure allows
determination of the shower direction, the core location and the height of the
shower maximum.  This technique has greatly improved the angular resolution
of imaging telescope arrays at the cost of reduced collection area. On the 
other hand the improvement of  angular resolution of a
wavefront sampling array results in a direct improvement in the flux
sensitivity since the already increased collection area of such arrays is
unaffected by the angle reconstruction. It may be noted here that a 
3-dimensional reconstruction of the shower is possible  for the a
non-imaging array by making use of the curvature of the \v Cerenkov light front 
\cite{VRC2k1c}.

\begin{table} \centering
\caption {Angular resolutions of various atmospheric \v Cerenkov experiments
in the world. The asterisk indicates that the quoted angular resolution refers 
to this primary energy. In the rest of the cases the quoted angular resolutions 
are not necessarily applicable at the $\gamma $-ray threshold energies shown in 
column 4. In the case of imaging telescopes it may be noted that the actual 
angular resolution could be worse than the pixel size quoted here.
\label{TAB:COMP} }
\vskip 0.25cm
\begin{tabular}{|l|l|l|l|l|}
\hline
& Observatory & Angular			  & $\gamma $-ray Energy  & Reference \\
& Name & Resolution ($deg.$)         & Threshold ($GeV$)  & \\
\hline
Imaging &  Whipple &  0.14 ($\in 68\%$) &   500   & \cite{fkr98} \\
\cline{2-5}
\v Cerenkov &  CAT  &   0.11 &   250   & \cite{pir96} \\
\cline{2-5}
Telescopes &  HEGRA(CT1) &  0.14 ($\in 68\%$) & 1700   & \cite{kra97} \\
\cline{2-5}
&  CANGAROO & 0.12  & 1000    & \cite{tde99}\\
\cline{2-5}
&  TACTIC   & 0.31  & 700 $\pm$ 200   & \cite{sbs97} \\
\cline{2-5}
& SHALON  & 0.4 & $1000^*$ &\cite{saa01}\\
& ALATOO I or II        &     &      & \\
\hline
 Stereoscopic &  HEGRA &  0.1  &  500   & \cite{goe01,dau97} \\
\cline{2-5}
 System & SHALON & 0.1 & $1000^*$ &\cite{saa01}\\
& ALATOO &&&\\
\hline
Non-Imaging & THEMISTOCLE & 0.63 ($\in 75\%$) & 3000 & \cite{bbd93}\\
\v Cerenkov &  STACEE  & 0.25 ($\in 68\%$) & 190 $\pm$ 60   & \cite{obb01} \\
\cline{2-5}
Telescope &  CELESTE &  0.2  & 60 $\pm$ 20   & \cite{dnm02} \\
\cline{2-5}
Arrays &  GRAAL  &  0.7 ($\in 63\%$) & 250 $\pm$ 110   & \cite{abb01} \\
\cline{2-5}
&  PACT  &  $2.4^\prime$ ($\in 68\%$)  & $3000^*$   &  \\
\cline{2-5}
& AIROBICC & $0.29\pm0.05$ ($\in 68\%$) & 30,000 & \cite{kmp95} \\
\hline
EAS & MILAGRO & 1.0   & 1000   & \cite{atk00}  \\
\cline{2-5}
    & TIBET III & 0.87 ($\in 50\%$) & 3000  & \cite{ame01}  \\
\hline       
Proposed &  VERITAS & $2.4^\prime$($\in 68\%$) &  $1000^*$  & \cite{bil99}  \\
\cline{2-5}
Imaging &  MAGIC &  0.1, 0.2 ($\in 68\%$) &   10   & \cite{bar98} \\
\cline{2-5}
Arrays &  HESS  &  0.1 ($\in 68\%$)  & 40   & \cite{aha97} \\
\cline{2-5}
&CANGAROO III & 0.1 &100 & \cite{mori99}\\
\hline
\end{tabular}
\end{table}

From the results of the present study it is seen that two independent 
measurements of the angular resolutions are $\sim 0^\circ .24$ and
$0^\circ .06$
resulting from the timing measurements of telescopes and mirrors of the entire
array respectively for similar energy events. The essential difference
between the two measurements is an 
increase in the number of degrees of freedom by a factor of 
$\sim~2.5$ in the latter case.  
From this consideration the expected angular resolution for individual
mirrors is $0.24/\sqrt{2.5} = 0^\circ .15$ which is almost twice the 
value of $0^\circ.06$ mentioned above. These are consistent since the timing 
uncertainty for {\it royal sums} is higher by the same factor.

\begin{figure}
\centerline{\psfig{file=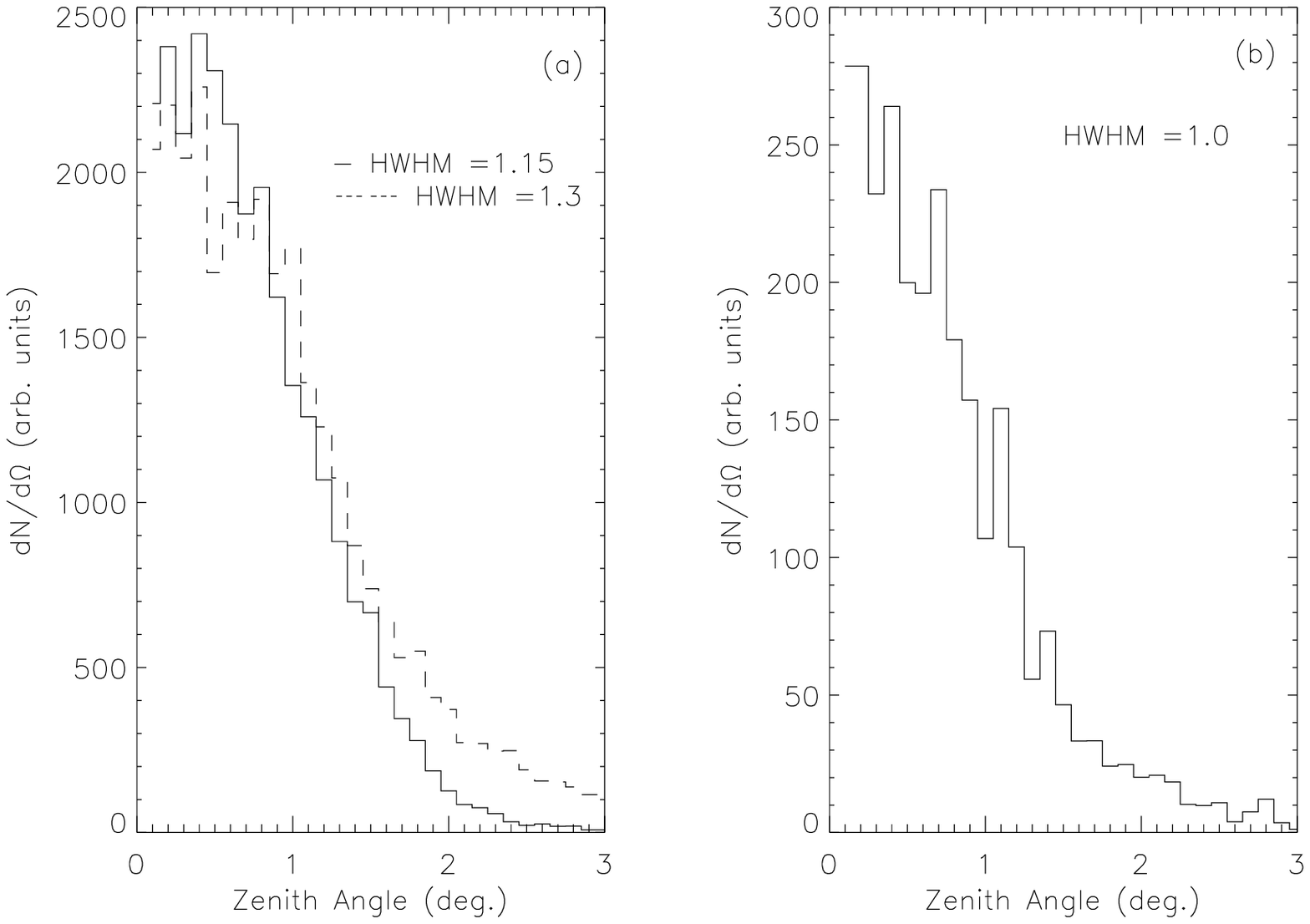,height=10cm}}
\caption{A differential plot of the number of events per solid angle as a 
function of zenith angle using estimates from  (a) {\it royal
sum} timing signals and (b) individual mirror timing signals.
The solid line histogram in (a) corresponds to {\it royal sum} events detected
simultaneously in all the 4 sectors while that in dotted line corresponds to 
{\it royal sum} events detected in sector 3 only. (See text for details.)
\label{Fig:dndom}}
\end{figure}

Figure \ref{Fig:dndom} shows differential plots of the number of events
per solid angle ${{dN}\over {d\omega}}$ as a function of the zenith angle. 
The panel $(a)$ shows plots 
for events whose arrival directions have been estimated using timing signals
from {\it royal sums} while the panel $(b)$ shows the same for individual 
mirror timing information. The two curves shown in panel $(a)$
correspond to events detected simultaneously in all the
4 sectors (solid line) and in one sector(\# 3) only (dashed line). 
This provides a method of measuring the effective 
opening angle of the array \cite{gmp88}.  From these plots it can
be seen that
the array field of view is around $2^\circ .3$. This is significantly 
lower than the geometric opening angle of $\sim 2^\circ .9$. It may be noted 
that the
effective field of view of the array is expected to be lower than the 
geometric value
primarily because the event trigger efficiency falls with increasing core 
distance and zenith angle. The other possible factors which could potentially 
reduce the
effective field of view like the residual alignment errors and non-ideal optical
quality of the mirrors are relatively less significant. The effective FWHM of
the array ($\sim 2^\circ .3$) for which the mean telescope
separation is $\sim 70~m$ is lower than that for a sector ($\sim 2^\circ .6$) 
for which the mean telescope separation is $\sim  32~m$ supporting
the above argument.

The data used for estimating the angular resolution of PACT reported here
consists of cosmic rays. It is well known that the protons exhibit a
larger (by almost a factor of 4) intrinsic timing jitter relative 
to $\gamma $-ray primaries \cite{VRC2k1}. While the contribution of systematic
effects to the timing resolution (which is estimated to be small) is independent
of the primary species, the decreased intrinsic timing jitter is expected to 
improve the angular resolution for $\gamma $-ray primaries significantly. 

Table \ref{TAB:COMP} summarizes the angular resolution of imaging as well as
non-imaging atmospheric \v Cerenkov telescopes currently in operation and
under construction. It can be seen that the angular resolutions of most 
of the single imaging telescopes are modest, in the range 
$0^\circ.1 - 0^\circ .7$. 
Wavefront sampling experiments using the modified solar concentrator arrays like
the STACEE, CELESTE, GRAAL $etc.$, will not be able to exploit their
angle reconstruction technique to reject a significant cosmic ray background
primarily because of their rather small field of view, in the 
range $0^\circ.24 - 0^\circ .7$ (FWHM) which are comparable to their angular
resolutions. PACT, on the other hand,  has the best angular 
resolution among the non-imaging experiments which can be used to enhance the
signal to noise ratio significantly. PACT has been able to 
achieve this because of multiple fast timing measurements at each telescope.
This has two-fold advantages: firstly, it provides an increased number of
degrees
of freedom which in turn improves the accuracy of angle reconstruction and 
secondly, it 
provides a means of computing the \v Cerenkov photon arrival time dispersion
at each telescope. The timing jitter is a species sensitive parameter
and hence will enable us reject a significant fraction of on-axis hadronic
showers.  The signal to noise
ratio could also be improved by using the normalized $\chi ^2$
values resulting from a spherical fit to the timing information of a shower.
This is expected to be larger for hadron initiated showers \cite{wbb00}.

The future imaging telescope arrays are expected to achieve 
unprecedented angular resolution by employing the stereoscopic technique. 
They are able to reconstruct the shower in 3-dimensions using the multiple
images in 3 or more imaging telescopes. From table \ref{TAB:COMP} it can be
seen that both the HEGRA and the SHALON-ALATOO arrays are able to improve 
the angular resolution of their single imaging telescopes by a factor of 
$\sim 4$ by using 2 or more of them in stereoscopic mode. It may be mentioned
here that the quoted angular resolutions for the future projects listed in
the table are from simulation results for $\gamma $-ray primaries while the rest
are derived from measurements from cosmic ray events.

It is well known that the \v Cerenkov light front has a curvature. It was seen 
that when two well separated \v Cerenkov telescopes were tilted towards each
other by about a degree the coincidence rate increased and also reduced the
spread in the time separation between them \cite{tnc68,grt85}. This indicated, 
as claimed by the authors, the
presence of curvature in the photon front. More recently, it was shown that
the radius of curvature of the front is equal to the height of the shower
maximum from the observation altitude \cite{VRC99}. Hence it is clear that 
a plane front approximation of the \v Cerenkov light front will introduce
a systematic error in the arrival angle reconstruction.
The large separation of telescopes, to some extent, offsets the worsening of
resolution due to the curvature of the light front.  It may also be 
pointed out that the effect of curvature on the angular resolution is more 
significant for near vertical 
showers.  In the case of inclined showers, the arrival time differences 
at spatially separated telescopes due to shower axis inclination
far exceed the differences arising out of the shower front curvature.
However it has been argued that the angular resolution of given array will
improve if one corrects for the wavefront curvature \cite{agg91}.
The details of the systematic effects due to the plane front approximation
and the improvement in the accuracy of estimated arrival directions when the
curvature of the shower front is taken into account are currently under
investigation. A paper based on the results of these simulation studies is 
under preparation.

\section{Summary}

  A detailed analysis of the angular resolution of PACT 
using data collected with telescopes pointing to zenith is presented.
The improvement in the angular resolution with larger separation between
detectors ({\it D}) and with increase in the number of degrees of freedom
({\it n}) has been verified.  The angular resolution $\sigma _\psi$, is found 
to improve
according to the relation:
   $${\sigma_\psi} \propto  {{1} \over {D \sqrt{n}}}$$

There are two types of angular resolutions that could be defined for PACT. 
Firstly, the fast timing information from the 25 telescopes spread in an area 
of about $80~m\times 100~m$ yield an angular resolution of 
$\sim 0^\circ .2$. Secondly the timing information from the mirrors
constituting the 25 telescopes yield the best angular resolution of
$\sim 2.4^\prime ~$ at a $\gamma $-ray energy of around 3 $TeV$. This is the 
best 
angular resolution achieved for any ground based atmospheric telescope
system which will probably be superseded by the proposed VERITAS or HESS
imaging telescope array.

The angular resolution for $\gamma $-ray primaries could be significantly
better than what is presented here, which is based on cosmic ray primaries, 
because
of two main reasons. Firstly, photon arrival time jitter for $\gamma $-ray 
primaries is far less than that for charged cosmic rays and secondly the 
radius of curvature for $\gamma $-ray primaries is more than that for cosmic
rays which consequently reduces the effects of the shower front curvature
for the former.

\ack{We would like to acknowledge the hardwork and dedication of our colleagues
who helped us in the installation of PACT, its instrumentation and the 
subsequent observations: Messrs S. S. Upadhya, K. Gothe, B. L. Venkateshmurthy,
B. K. Nagesh, K. K. Rao, A. J. Stanislaus, S. K. Rao, P. V. Sudershanan, P. N. Purohit, 
S. K. Sharma, M. S. Pose, A. I. D'Souza, S. R. Joshi. We would like to thank 
the anonymous referee for a careful, patient and thorough job of refereeing this
paper.} 

\end{document}